\documentclass[]{aa}  
\usepackage{verbatim}
\usepackage{color}
\begin{document}

\title{Anharmonicity and the infrared emission spectrum of highly excited PAHs}

   \author{Tao Chen
          \inst{1,2}\fnmsep\thanks{Email: chen@strw.leidenuniv.nl}
          \and
          Cameron Mackie\inst{1}
          \and
          Alessandra Candian\inst{1}
          \and
          Timothy J. Lee\inst{3}
          \and
          Alexander G. G. M. Tielens\inst{1}
          }

   \institute{Leiden University, Leiden Observatory, Niels Bohrweg 2, NL-2333 CA Leiden, Netherlands\\
         \and
         School of Engineering Sciences in Chemistry, Biotechnology and Health, Department of Theoretical Chemistry \& Biology, Royal Institute of Technology, 10691, Stockholm, Sweden\\
         \and
         NASA Ames Research Center, Moffett Field, California 94035-1000, USA
          }

\abstract
  {} 
  {Infrared (IR) spectroscopy is a powerful tool to study molecules in space. A key issue in such analyses is understanding the effect that temperature and anharmonicity have on different vibrational bands, and thus interpreting the IR spectra for molecules under various conditions.} 
  {We combined second order vibrational perturbation theory and the Wang-Landau random walk technique to produce accurate IR spectra of highly excited PAHs. We fully incorporated anharmonic effects, such as resonances, overtones, combination bands, and temperature effects.}
{The results are validated against experimental results for the pyrene molecule (C$_{16}$H$_{10}$). In terms of positions, widths, and relative intensities of the vibrational bands, our calculated spectra are in excellent agreement with gas-phase experimental data.}
   {}
   \keywords{anharmonicity --
             temperature effects --
             aromatic infrared bands --
             infrared emission spectra --
             polycyclic aromatic hydrocarbons
            }

   \maketitle

\section{Introduction}
Polycyclic aromatic hydrocarbons (PAHs) are a class of organic molecules that are generally thought to be responsible for aromatic infrared bands (AIBs). This family of discrete emission features has been recorded in the mid-infrared (3 \textendash \, 20 $\mu$m; mid-IR) from a wide variety of astronomical objects in the Milky Way and galaxies at least to $z=4$ \citep{sajina2007spitzer, riechers2014polycyclic}. These features are very strong, accounting for 20 \textendash \, 30\% of the infrared (IR) galactic emission and up to 10\% of the total output in star-forming galaxies \citep{tielens2008}.

PAH molecules are excited by the absorption of a visible/UV photon and undergo internal conversion, where the excess energy is converted from electronic to nuclear degrees of freedom, while they relax to the electronic ground state. Subsequently, the energy is redistributed among the vibrational degrees of freedom (i.e., intramolecular vibrational redistribution; IVR) and the vibrationally excited molecules cool down through emitting IR photons \citep{allamandola1985polycyclic, allamandola1989interstellar}.

Because of the highly challenging nature of the experiments, only a few emission spectra of highly excited, gaseous PAHs have been recorded using a very sensitive IR detector \citep{cook1998simulated, wagner2000peripherally}. These spectra show that band positions are shifted with respect to those of low-temperature absorption spectra and that the shift, as well as the band width, depends on the internal excitation of the molecule. These effects are clear signs of the anharmonicity of the potential energy surface (PES) \citep{cook1998simulated}.

From the theoretical point of view, several methods can be used to calculate IR vibrational spectra that take excitation and anharmonic effects into account. Generally, the anharmonicities, basis set incompleteness, and approximations in the treatment of electron correlation can be corrected using empirical scaling factors applied on the harmonic frequencies obtained from a static quantum chemical calculation and this has been widely used in the astrophysical literature \citep{langhoff1996theoretical, bauschlicher2008infrared}. Molecular dynamics (MD) simulations can also well describe the excitations and anharmonicities. These simulations provide the IR spectrum upon Fourier transformation of the autocorrelation of electric dipole moment. As no assumptions about the shape of the PES are made, MD intrinsically accounts for resonances, rovibrational couplings, and temperature effects \citep{schultheis2008, van2002vibrational, kumar2006understanding, estacio2008born}. 

Alternatively, temperature-dependent IR spectra can be calculated by a time-independent or static method, which is incorporated through inverse Laplace transformation of quantum partition functions \citep{romanini1993numerical}. The quantum partition functions are related to the microcanonical density of vibrational states, which are difficult to calculate for polyatomic molecules; the problem has only been satisfactorily solved for separable oscillators \citep{stein1973accurate}. The vibrational density of states (DoS) can also be estimated by direct Monte Carlo (MC) integration \citep{noid1980properties, farantos1982monte, barker1987sums}. However, such methods become highly inefficient for large systems \citep{hupper1999numerical}. An improved MC method, called the Wang-Landau method, has been introduced for handling large-dimensional space problems \citep{wang2001efficient}. It has been shown that this method accurately reproduces vibrational DoS for large nonseparable systems \citep{basire2009, calvo2010finite}. 

However, resonances and combination bands were not considered in this method. Recent works reveal that resonances and combination bands are important for comparison with high-resolution experimental IR spectra of PAHs, especially for the CH-stretching modes around 3000 cm$^{-1}$ \citep{mackie2015anharmonic, maltseva2015high}. In this work, we apply the Wang-Landau method to calculate the IR emission spectrum of highly vibrationally excited pyrene, taking the effects of anharmonicity and resonances into account. This study is based upon the theoretical analysis of recent high-resolution spectra, which is calculated using second-order Vibrational Perturbations Theory (VPT2) calculations \citep{mackie2015anharmonic}.  

\section{Computational details}
The IR spectra at a given internal energy were calculated by first computing the vibrational DoS using the Wang-Landau method \citep{basire2009}. We then accumulated the two-dimensional histogram of the intensity of transitions; finally, we converted the accumulated absorption histogram to microcanonical absorption spectrum at a fixed internal energy. For comparison to the NIST spectra \citep{linstrom2001nist}, obtained in thermal equilibrium, we derive the absorption intensity at finite temperature by a Laplace transformation. 

The vibrational DoS are calculated by a random walk in energy space \citep{basire2009}
\begin{equation}
E(n) = \sum_i h\omega_i (n_i + \frac{1}{2}) + \sum_{i \le j}\chi_{ij}(n_i + \frac{1}{2})(n_j + \frac{1}{2}),
\end{equation}
where {n} $\equiv$ (n$_1$, n$_2$, ...), are the quantum numbers representing the states of each vibrational energy level. The harmonic vibrational frequencies $\omega_i$ and anharmonic couplings $\chi_{ij}$ are calculated using VPT2 based on a semidiagonal quartic force field computed with density functional theory (DFT) as implemented in the Gaussian 09 package \citep{frisch2009gaussian}. The hybrid density functional B3LYP method \citep{becke,lee1988development} with the N07D basis set \citep{barone2008accurate} are utilized for all the calculations. For the geometry optimization a very tight convergence criterion and a very fine grid (Int = 200 974) are used in the numerical integration \citep{cane2007anharmonic, bloino2015anharmonic}. The resonant states are evaluated through VPT2 approach, in which Fermi, Darling-Dennison, and vibrational Coriolis resonances are treated through a polyad approach \citep{martin1995anharmonic} as described previously \citep{mackie2015anharmonic}. At any point during the random walk, resonance effects on the vibrational frequencies have to be evaluated using the polyad as the excitation in the different modes changes.

The effects of internal excitation are included through a second MC simulation in the space of quantum numbers. In brief, during the random walk, the quantum numbers ($n$) of the different modes changes, which directly affects the excitation energy of each mode (cf. Eq. 1). At every step the effects of anharmonicity and resonances have to be evaluated in order to obtain a reasonable band width and position, and in the treatment of resonances only the diagonal elements of the resonance matrix have to be updated, which makes for a very efficient algorithm. The energy dependent spectra are accumulated during the second Wang-Landau walks following the previously calculated DoS. At each Wang-Landau step, an IR spectrum (produced by that particular random configuration of quanta) is added to that energy-dependent spectrum. The line positions are determined by Equation 1 and intensities by the following formula:

\begin{equation}
I(\omega,E) = (n_k+1)\sigma_{\omega_k} 
,\end{equation}
where I(v,E) is the intensity of vibrational mode v at energy E, $n_k$ is the number of quanta in mode k, and $\sigma_{\omega_k}$ is the absorption cross section of mode $\omega_k$ \citep{basire2009}. 

It is possible to include rotational excitation effects on the vibrational spectra as well \citep{basire2009}. However, as we are ultimately interested in the UV-pumped IR fluorescence of initially very cold PAH species in the interstellar medium (ISM)  and rotations are poorly coupled to the vibrational excitation \citep{ysard2010long}, we ignored rotational effects in this work.

\begin{figure}
\includegraphics[width=1.0\columnwidth]{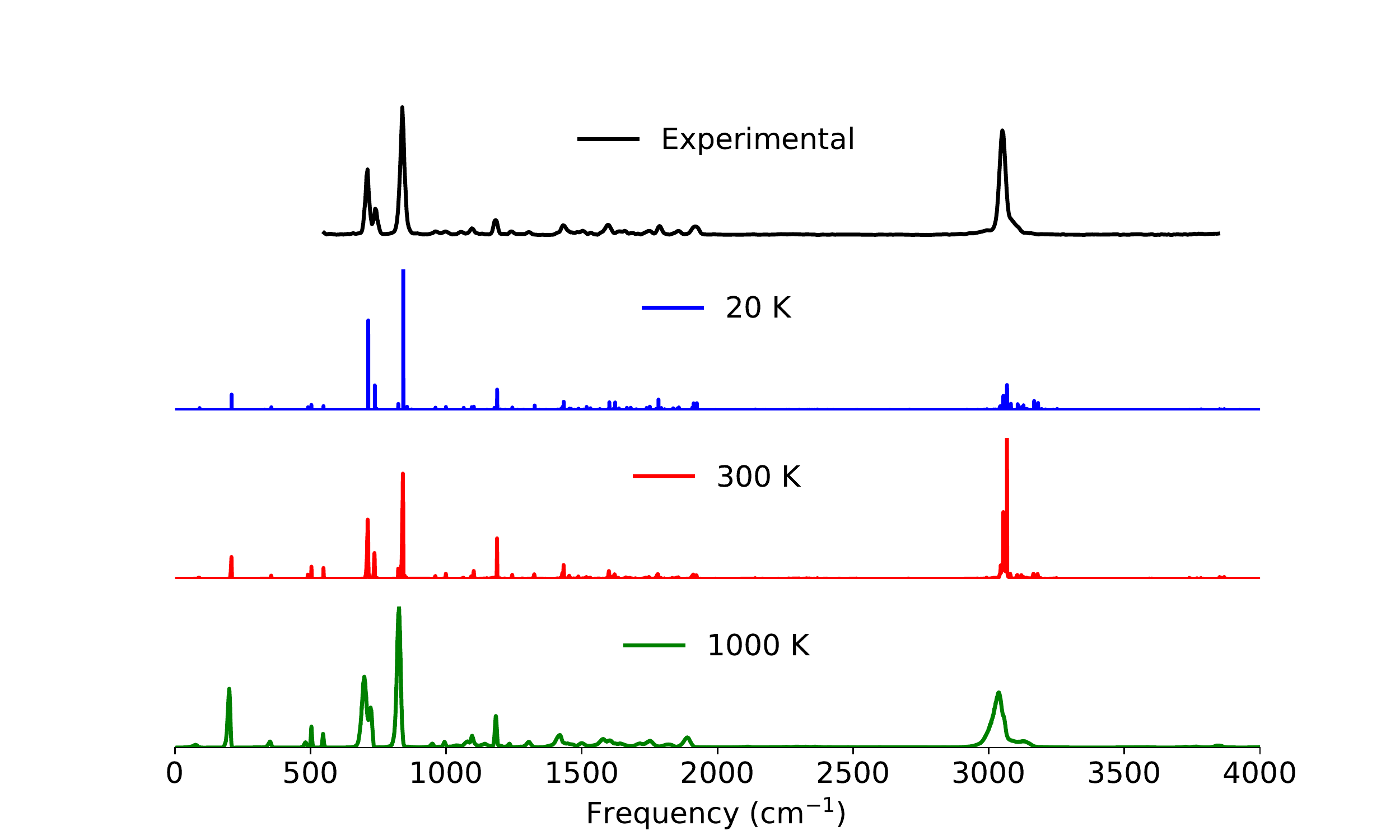}
\caption{Gas-phase experimental NIST IR spectrum of pyrene ($\sim$ 300 K) and comparison with calculated anharmonic spectra at temperatures 20 K, 300 K, and 1500 K. No artificial broadening is applied. The broadening and peak shift are a direct consequence of the anharmonic interaction.}
\label{fig:pyrene_nist}
\end{figure}

\begin{figure}
\includegraphics[width=1.0\columnwidth]{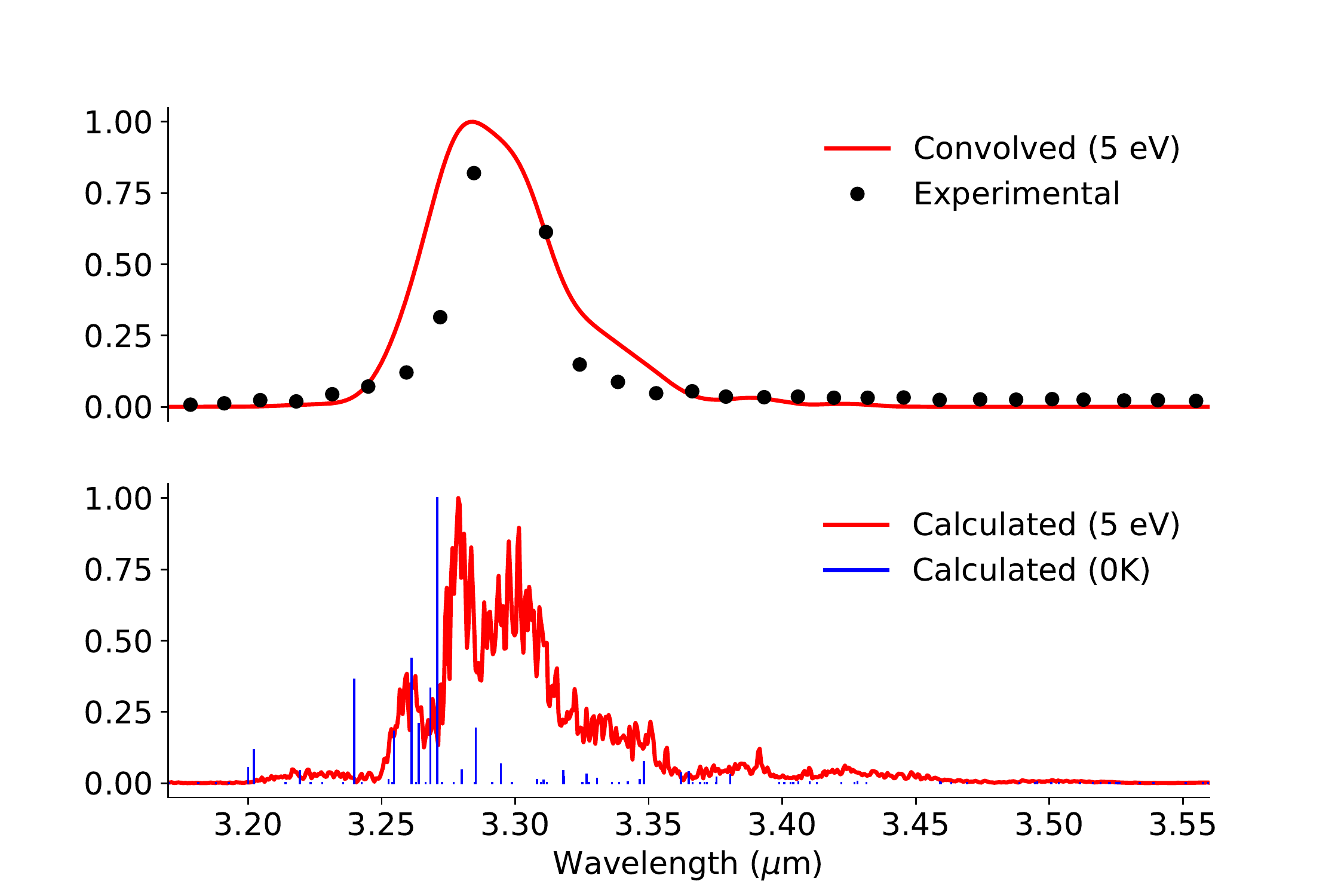}
\caption{Temperature effects on IR spectrum of pyrene and comparison with the experimental spectrum (black dots), which was measured using SPIRES \citep{wagner2000peripherally}. The red curves represent the calculated anharmonic emission spectrum after absorbing a 5 eV photon. In order to compare the band profiles with the experimental data, the calculated spectrum in the top panel is convolved with Gaussian functions. The blue lines represent the fundamental and combination bands calculated at 0 K, i.e., without temperature effects considered in the calculations.}
\label{fig:pyrene_wagner}
\end{figure}

\begin{figure}
\includegraphics[width=1.0\columnwidth]{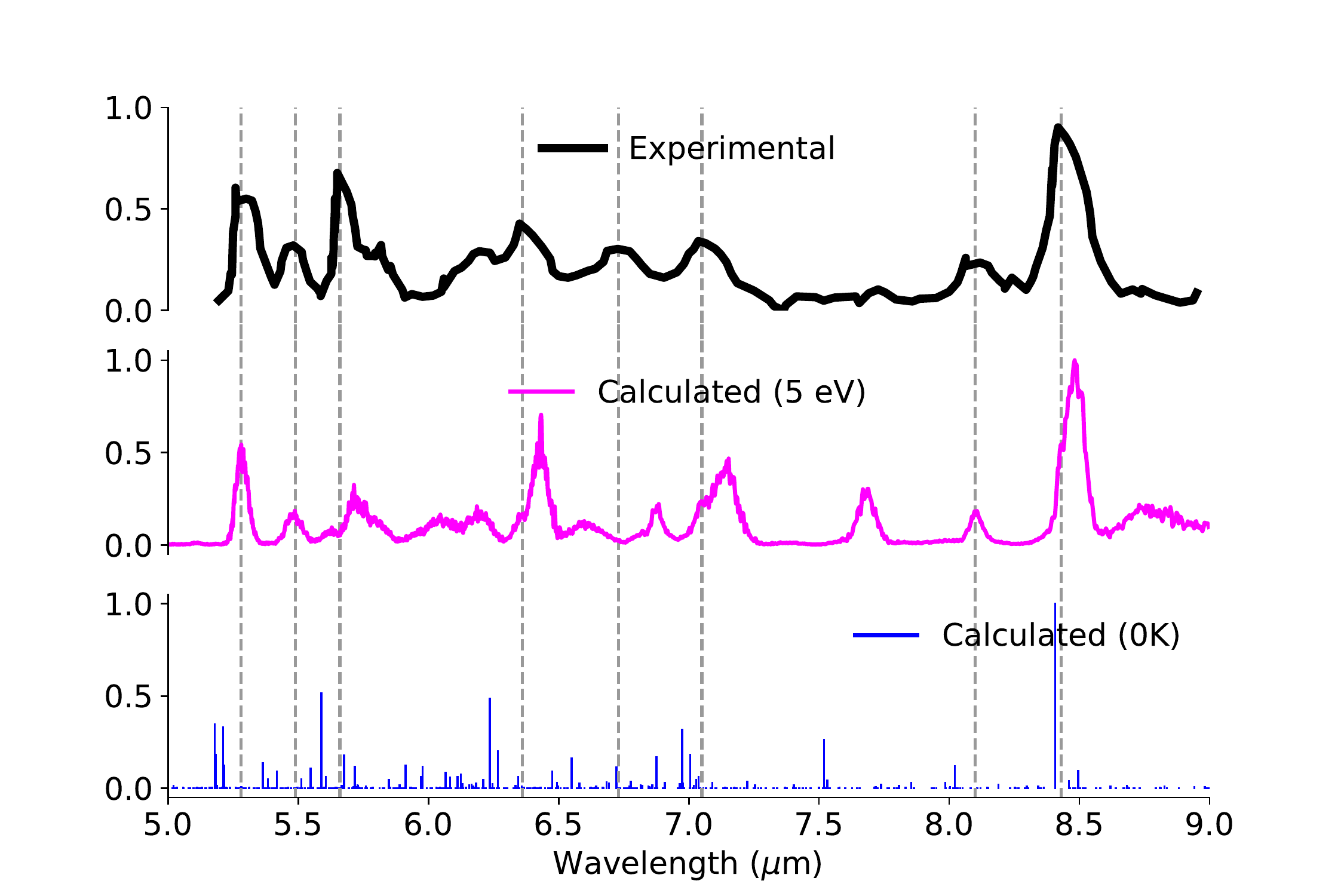}
\caption{Temperature effects on IR spectrum of pyrene in 5 \textendash \, 9 $\mu$m region and comparison with experimental data (blue curve in the top panel) \citep{cook1996infrared}. The red curve represents the calculated anharmonic emission spectrum after absorbing a 5 eV photon. The blue sticks are the fundamental and combination bands calculated at 0 K, i.e., without temperature effects considered in the calculations. The gray dashed lines show prominent bands observed on the experimental spectrum.}
\label{fig:pyrene_cook5-9}
\end{figure}

\begin{figure}
\includegraphics[width=1.0\columnwidth]{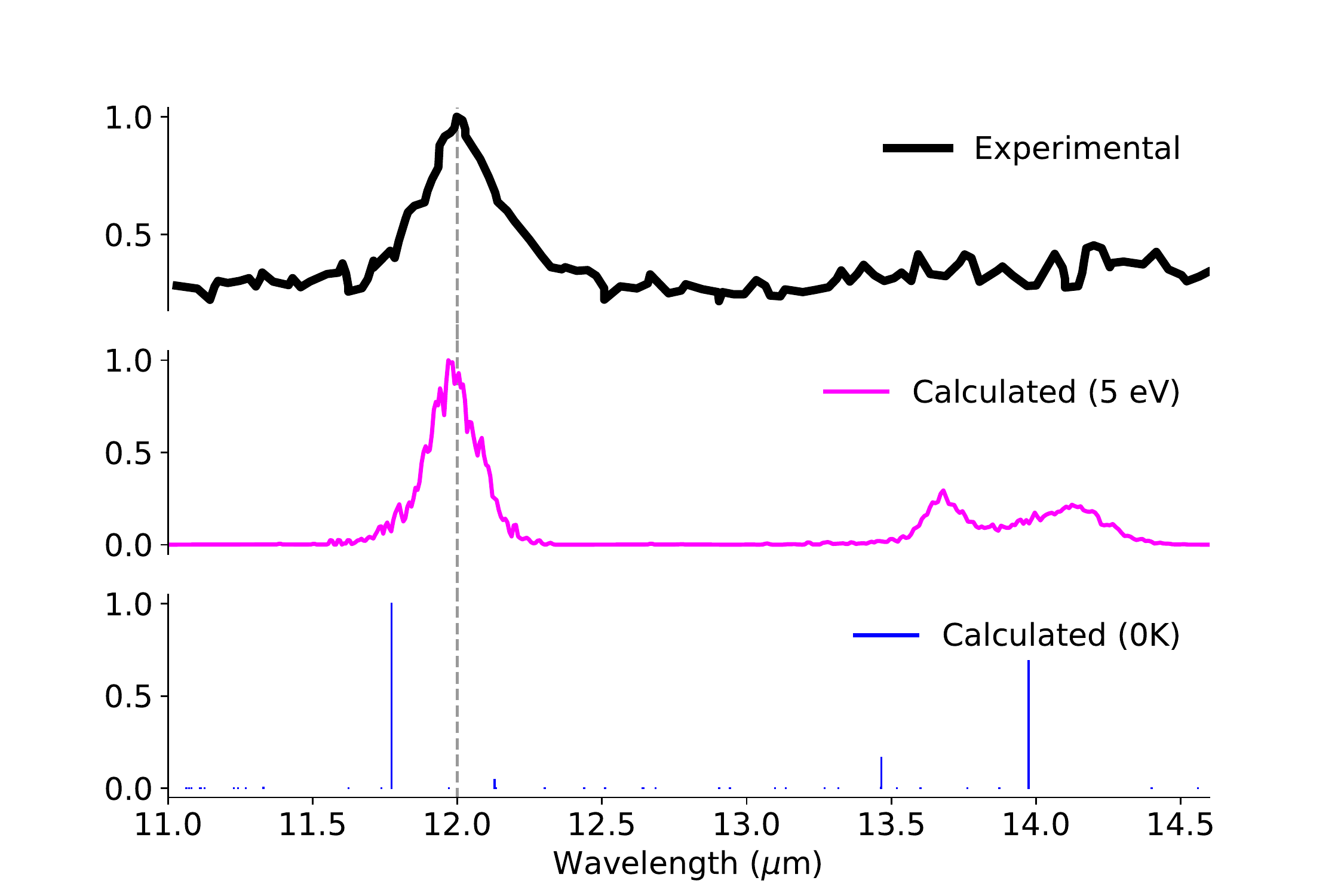}
\caption{Temperature effects on IR spectrum of pyrene in 12 $\mu$m region and comparison with experimental data (blue curve in the top panel) \citep{cook1996infrared}. The red curve represents the calculated anharmonic emission spectrum after absorbing a 5 eV photon. The blue sticks are the fundamental and combination bands calculated at 0 K, i.e., without temperature effects considered in the calculations. The gray dashed line shows the main band observed on the experimental spectrum.}
\label{fig:pyrene_cook12}
\end{figure}

\section{Results and discussion}
The method has been validated through comparison with experimental spectra downloaded from the NIST Chemistry WebBook \citep{linstrom2001nist}. Figure~\ref{fig:pyrene_nist} shows the temperature effects on the IR spectrum of pyrene (C$_{16}$H$_{10}$) and comparison with the NIST data. The NIST spectrum is recorded at room temperature, i.e., $\sim$ 300K. The calculation generates reasonable spectra as shown in Figure~\ref{fig:pyrene_nist}. The broadening and peak shift due to the increase of temperature are a direct consequence (in terms of peak position and relative intensity) of the anharmonic interaction, i.e., no artificial broadening or frequency correction factor is applied. Owing to the lack of rotational effects in the calculation, the calculated spectra are narrower than the NIST data. 

Figure~\ref{fig:pyrene_wagner} compares the IR emission spectra of the gas-phase UV laser-excited pyrene in the C-H stretching region (3.3 $\mu$m). The experimental spectrum was measured using the Berkeley Single Photon InfraRed Emission Spectrometer (SPIRES) \citep{schlemmer1994, wagner2000peripherally}. A plume of pyrene is vaporized into the gas phase using a 258 nm laser. The absorption of a laser photon also provides an internal energy of 5 eV into the pyrene molecule. The spectrum is integrated over some 100 microseconds as the plume expands. The spectra of several laser shots obtained over 10-100 seconds\, are summed to increase the signal to noise. As radiative relaxation takes about 1 s, each individual spectrum corresponds to pyrene with an internal energy of 5 eV. This elegant experiment is well designed to simulate the environment of the ISM as pyrene is highly vibrationally excited but rotationally cold. This also provides an excellent platform to validate our model. Comparing the spectra calculated at 5 eV internal energy with those at 0 K shows the effect of internal excitation on the anharmonic interaction, which leads to a pronounced redshift and broadening of the individual features (Figure~\ref{fig:pyrene_wagner}). A vibrational temperature can be estimated using the SPIRES frequencies in conjunction with the redshift measurements, which yields a temperature of 1019 K for pyrene \citep{joblin1994infrared, cook1996infrared}. Much of the individual structure blends into several broad features. At the resolution of the experimental spectra ($\sim$ 100), these broad features blend together even further. The resulting calculated spectrum is in good agreement with the measured spectrum in peak position, overall profile, and width.

Figure~\ref{fig:pyrene_cook5-9} compares the emission spectrum of pyrene in the 5 \textendash \, 9 $\mu$m region as measured by SPIRES \citep{cook1996infrared}. Again, there are no free parameters in this analysis as anharmonic interaction dictates the frequency shift, broadening, and intensity of the bands. Over this wavelength regime, calculated peak positions are in reasonable agreement with the experiments but the calculated bandwidth seems somewhat narrower than the experiments indicate. We note that the bands shortward of 6 $\mu$m are combination bands, which in the double harmonic approximation have no intrinsic intensity. In anharmonic calculations, these bands obtain IR intensity through inclusion of an anharmonic dipole surface. The good agreement in calculated (relative) strength for all bands over this full wavelength region is therefore very impressive.  

The 11 \textendash \, 15 $\mu$m region contains the out-of-plane bending modes. Pyrene has duo and trio H's \footnote{Duo: two adjacent C atoms, each with an H atom. Trio: three adjacent C atoms, each with an H atom.}, This results in one strong band at $\sim$11.7 $\mu$m where both the duos and trios move in unison out of the plane. The weaker longer wavelength bands show combined activity in the CH out-of-plane bending and CCC skeletal motions. Again, the bands shift and broaden significantly upon excitation (Figure~\ref{fig:pyrene_cook12}). The calculated peak position and bandwidth of the out-of-plane bending mode are in good agreement with the experiment. Surprisingly, the longer wavelength modes are not present in the experiments, perhaps reflecting limited signal to noise in this region.

Overall, the agreement between the calculations and experiments is good. Peak positions agree typically to better than 1.5\%. The agreement in width is somewhat worse ($\sim$ 30\%). This agreement is particularly impressive given that there are no adjustable parameters. Both frequency shifts, broadening, and relative strength are fully prescribed by the anharmonic and resonance interactions between the modes calculated using VPT2/DFT, and the other controlling parameter, the internal energy, is fixed by the laser energy. The remaining small differences in peak position may reflect small inaccuracies in the calculated anharmonic coefficients as the effects of these get amplified upon excitation. In the 3 $\mu$m region of the spectrum, the effects of resonances are strongest. In this wavelength region, differences between calculated and measured spectra may reflect the importance of triple combination bands as experiment and theory indicate that these are important in understanding the number of bands and their relative strength in low temperature absorption spectra \citep{maltseva2015high}. Further improvement in the theory will have to await development of efficient methods to calculate the effects of triple combination bands on the spectra of highly excited molecules and/or a further refinement of the calculated anharmonic interaction coefficients. 

With the launch of the James Webb Space Telescope, moderate resolution (R $\sim$ 3000) mid-IR (3 \textendash \, 28 $\mu$m) spectroscopy of a wide range of astronomical sources is coming into reach. The results shown in this paper go a long way toward validating astronomical PAH emission models. If this model can be extended in an efficient way to larger PAHs and derivatives, this carries the promise of an in-depth analysis of the astronomical data and a deep understanding of the interstellar PAH family. 

\section{Conclusions}
In this work, we calculated the anharmonic IR absorption and emission spectra of pyrene. The VPT2 and Wang-Landau method are combined in order to account for temperature effects on the molecule. The results are compared with experimental gas-phase IR spectra. The band position, band profile, and intensities of the experimental spectra are accurately reproduced by the model. The results demonstrate that anharmonicity and temperature effects are necessary for understanding the PAH spectra. The model can be used as a powerful tool for studying the fine structure of IR spectra, which could provide detailed information regarding the ecosystem of molecules in space. This model might also be used for species identification and for extrapolating to areas where experimental data are limited.

\begin{acknowledgements}
This work is supported by Swedish Research Council (Contract No. 2015-06501). The facility is supported by the Swedish National Infrastructure for Computing (Project No. SNIC 2018/3-30). The calculations were carried out on Kebnekaise and Abisko located at High Performance Computing Center North (HPC2N). 

We thank Julien Bloino for fruitful discussions and generously providing knowledge on the GVPT2 calculations. Special thanks to Ye Zhang and Andrey R. Maikov. AC acknowledges NWO for a VENI grant (639.041.543). TJL is supported by the National Aeronautics and Space Administration through the NASA Astrobiology Institute under Cooperative Agreement Notice NNH13ZDA017C issued through the Science Mission Directorate as well as support from the NASA 16-PDART16 2-0080 grant. We acknowledge the European Union (EU) and Horizon 2020 funding awarded under the Marie Sk\l{}odowska Curie action to the EUROPAH consortium, grant number 722346. Studies of interstellar PAHs at Leiden Observatory are supported through a Spinoza award.
\end{acknowledgements}

\end{document}